\documentclass[12pt]{article}
\usepackage[margin=1in]{geometry}

\usepackage{amsmath}
\usepackage{graphicx}

\newtheorem{Theorem}{Theorem}

\usepackage{graphicx}
\usepackage{multirow}
\usepackage{mdwlist}

\usepackage{threeparttable}

\usepackage{amsfonts}
\usepackage{amsmath}
\usepackage{hyperref}
\usepackage{float}
\usepackage{caption}
\usepackage{subcaption}

\begin{document}
\title{\bf Analysis of a Markovian Queuing Model for Autonomous Signal-Free Intersections}

\author{Hanwen Dai and Li Jin
\thanks{This work was in part supported by the C2SMART University Transportation Center, and the NYU Tandon School of Engineering.}
\thanks{The authors are with the Tandon School of Engineering, New York University, Brooklyn, New York, USA. 
H. Dai is also with the Department of Industrial Engineering, Tsinghua University, Beijing, China.
Emails: hd2145@nyu.edu, lijin@nyu.edu}%
}

\maketitle

\begin{abstract}
We consider a novel, analytical queuing model for vehicle coordination at signal-free intersections.
Vehicles arrive at an intersection according to Poisson processes, and the crossing times are constants dependent of vehicle types.
We use this model to quantitatively relate key operational parameters (vehicle speed/acceleration, inter-vehicle headway) to key performance metrics (throughput and delay) under the first-come-first-serve rule.
We use the Foster-Lyapunov drift condition to obtain stability criteria and an upper bound for average time delay.
Based on these results, we compare the efficiency of signal free intersections with conventional vehicles and with connected and autonomous vehicles. We also validate our results in Simulation of Urban Mobility (SUMO).
\end{abstract}


\section{Introduction}


Modern connected and autonomous vehicle (CAV) technology has a strong potential for improving urban traffic efficiency.
A connected and fully autonomous vehicle is capable of sensing the environment, communicating with other vehicles and infrastructures, and using the advanced control system to process the gathered information and decide its movement and navigation path \cite{guo2019, malikopoulos2014}.
In particular, the CAV technology enables high-speed signal-free intersections, where a centralized controller coordinates the movement of CAVs crossing an intersection with point-to-point guidance \cite{rios2015}.
An efficient coordinate algorithm of intersection control can reduce time delay and save energy for autonomous vehicles \cite{miculescu2019polling}.


In this paper, we study the performance of signal-free intersections in an analytical manner. We propose a Markovian queuing model for automated intersections. Our model is analogous to M/D/$\infty$ models but has two important and peculiar features for signal-free intersections.
First, we allow delayed parallel service.
Second, we impose a switching delay. Vehicles cross the intersection on a first-come-first-serve (FCFS) basis. Using the queuing model, we provide a system stability criteria and an upper bound of the average vehicle time delay. To validate our theoretical findings, we use Simulation of Urban Mobility (SUMO \cite{krajzewicz2010traffic}) to run experiments and make comparison. We find that our theoretical results match the simulation results well. 

\begin{figure}[htbp]
    \centering
    \includegraphics[width = 0.35\textwidth]{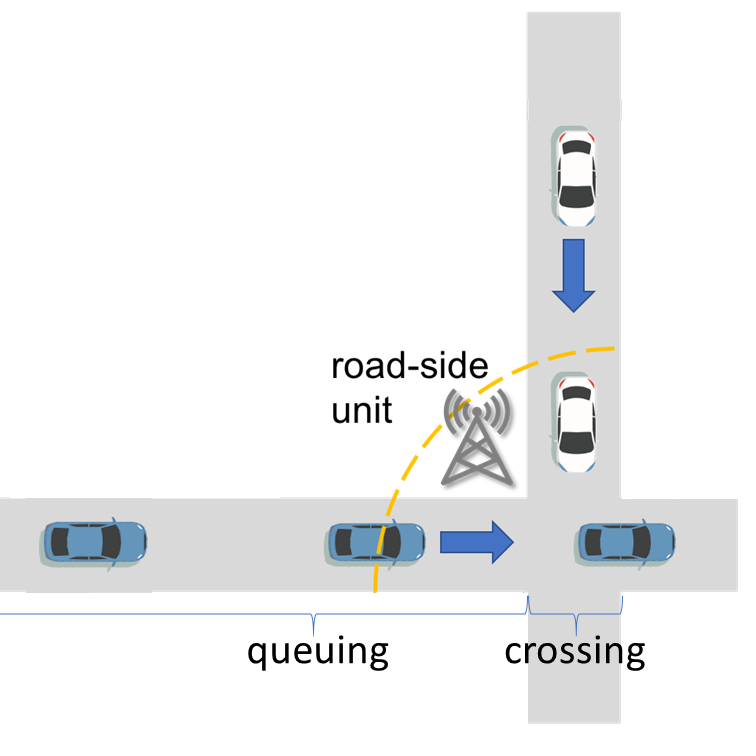}
    \caption{A signal-free intersection with CAVs.}
    \label{fig:intro}
\end{figure}




The merging of multiple traffic streams is a major component of traffic flow problems. Early in 1969, a unified approach was established to solve the problem of merging two or more strings of high-speed vehicles into a single lane~\cite{athans1969}; Dresner and Stone~\cite{dresner2004} proposed a reservation-based system for coordinating vehicles at intersections in 2004. A $n$-path merging problem can be regarded as a scheduling problem~\cite{colombo2015}, which involves determining the merging sequence and vehicle trajectory with collision avoidance.
Many researches use kinematic models that track vehicle position, velocity and acceleration/deceleration~\cite{colombo2015, kamal2013, ntousakis2016, rios2016, rios2015, zhang2017b, zhang2016, zhang2017a, zohdy2012}; at the more macroscopic level, Zhu and Ukkusuri proposed a lane based traffic flow model to reduce control complexity at intersections~\cite{zhu2015}. Our queuing model stands between the kinematic model and the traffic flow model; our model captures both microscopic vehicle heterogeneity/interaction and macroscopic performance metrics. Related to Miculescu and Karaman~\cite{miculescu2019polling}'s research work of extending the polling system to intersection control, our work gives the theoretical analysis of the two-queue polling system with constructing Lyapunov function and referring to Foster-Lyapunov criteria.

We formulate the intersection control problem as a continuous time two-queue system with Poisson arrival process, constant crossing time depending on the vehicle class, and a cooldown time between two consecutive vehicles. Unlike conventional queuing models, we define the system state as a tuple of the residual system time, the class, and the constant crossing time of the last entered vehicle.
Thus, our model is continuous-time and continuous-state, which cannot be analyzed using classical queuing-theoretic approaches.
To address this, we proceed with an infinitesimal generator for the queuing processes, which is derived based on the theory of piecewise-deterministic Markov processes \cite{davis1984piecewise}.


We analyze the system stability with the help of the Foster-Lyapunov drift condition \cite{meyn1993stability}. To achieve system stability, the proposed quadratic Lyapunov function should drift in the negative direction as the residual system time of the last entered vehicle increases, which comes to our stability criteria. The drift condition also leads to an upper bound of the average time delay.
We use the upper bound as a proxy for the exact delay, which is very hard to compute.


Finally, we run simulation by using SUMO to validate our theoretical findings. Since our queuing model makes some simplification and disregard some details, such as vehicle following model and vehicle acceleration profile. SUMO is able to provide a more realistic intersection environment and vehicle interaction. Given vehicle flow input and running the simulation, SUMO will return the average time delay to be compared with the theoretical results.

The main contributions of this paper are as follows. First, we develop an analytical queuing model integrating the advantages from macroscopic traffic flow model and microscopic vehicle dynamics model. Second, based on our model, we provide the system stability criteria and an upper bound of the average time delay.
Third, we validate our modeling and analysis approaches via microscopic simulation.

The rest of this paper is organized as follows. In section \ref{sec:model}, we formulate the signal-free intersection as a polling system with two queues. We provide theoretical analysis for system stability in section \ref{sec:stability}. In section \ref{sec:simulation}, we validate our theoretical findings through simulation experiments. We conclude the paper with remarks in section \ref{sec:conclusion}. 
\section{Modeling and formulation}
\label{sec:model}

We model a two-direction intersection as a system of two queues.
We label the classes of vehicles in these two directions as 1 and 2.
Vehicles join each queue as Poisson processes of rate $\lambda_1$ and $\lambda_2$, respectively.
\begin{figure}[htbp]
    \centering
    \includegraphics[width = 0.35\textwidth]{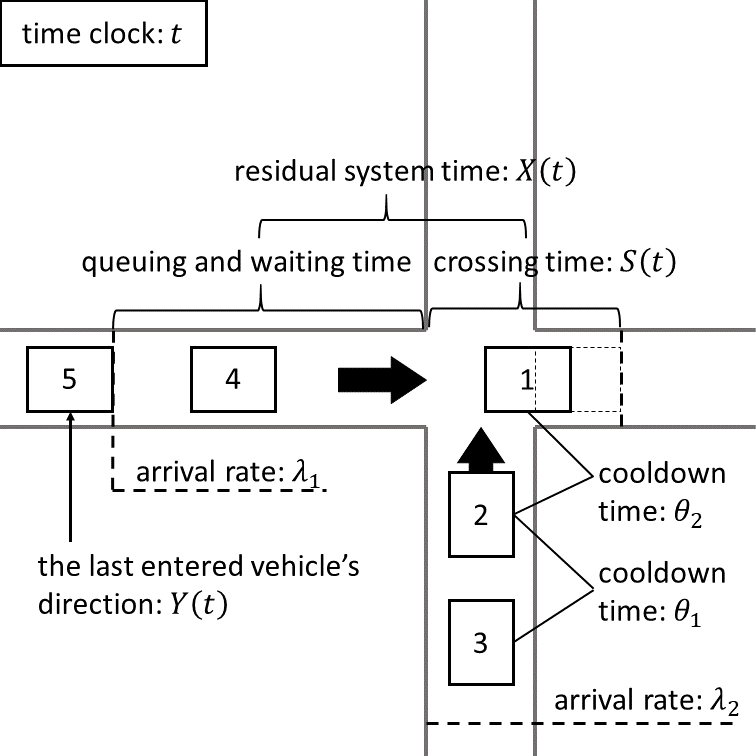}
    \caption{The queuing model for a two-way intersection.}
    \label{fig:intersection}
\end{figure}
Every vehicle will experience two stages as it go through the intersection, viz. queuing and crossing; see Fig.~\ref{fig:intersection}.
The queuing occurs outside the crossing zone, where vehicles queue and wait for discharge instruction from the rode-side unit (RSU).
After the RSU instructs a vehicle to enter the crossing zone, the vehicle will traverse the crossing zone with a time $S$ and at speed $u$.
To capture the heterogeneity in traffic mixture, we assume that $S$ are independent and identically distributed (IID) random variables with a discrete probability mass function (PMF) $\{p_s;s\in\mathcal S\}$, where $\mathcal S$ is the set of vehicle type-specific (e.g. cars, trucks, buses, etc.) crossing times.

We model the crossing zone as $\infty$ servers with vehicle type-dependent crossing (service) times. Note that, although the crossing time is heterogeneous due to the heterogeneity of vehicle types, it is known to the RSU before the crossing starts. However, these servers are not independent, and we can not let infinite vehicles enter the crossing zone; instead, the system of servers is subject to a \emph{cooldown} time, and the number of vehicles crossing the intersection will be constrained by the cooldown time.
The cooldown time captures the safety constraint on inter-vehicle headways.
That is, if one server starts serving a vehicle (the leading vehicle) at time $t$, then any other server cannot start serving the next vehicle (the following vehicle) until the time $t+c$; the cooldown time $c$ is equal to $\theta_1$ if the leading and following vehicles are from the same direction and is equal to $\theta_2$ if they are from distinct directions.
In general, $\theta_2\ge\theta_1$; this is similar to the switch-over time in \cite{miculescu2019polling}. In this paper, we assume that the cooldown time is smaller than the minimal crossing time.

Note that the queuing model as described above would be not Markovian if we were to use the number of vehicles as the state variable. The reason is that the crossing time, which resembles the service time in classical queuing models, is not exponentially distributed. To resolve this, we consider the state $(X(t),Y(t),S(t))$, where $X(t)\in\mathbb R_{\ge0}$ is the  residual system time for the vehicle that last entered the system up to time $t$, and $Y(t)\in\{1,2\}$ is the class of the vehicle that last entered the system up to time $t$ and $S(t)\in \mathcal S$ is the crossing time of the vehicle that last entered the system up to time $t$.

We assume that vehicles are scheduled to cross the intersection in a first-come-first-serve (FCFS) manner. Note that our modeling and analysis approach apply to alternative schedule schemes as well.
When a vehicle arrives at time $t$, $Y(t)$ and $S(t)$ are updated according to the class of this vehicle. Then, $X(t)$ is updated as follows:
{\footnotesize\begin{align*}
    & X(t)=\\
    & \begin{cases}
    X(t_-)+\theta_1 + S(t) - S(t_-) & Y(t_-)=Y(t), X(t_-) \ge S(t_-)-\theta_1,\\
    X(t_-)+\theta_2 + S(t) - S(t_-) & Y(t_-)\ne Y(t), X(t_-) \ge S(t_-)-\theta_2,\\
    S(t) & \mbox{otherwise,}
    \end{cases}
\end{align*}}%
where $S$ is a random variable with PMF $p_S(s)$.
Between two vehicle arrivals, $X(t)$ evolves as follows:
\begin{align*}
    \frac d{dt}X(t)=\begin{cases}
    0 & X(t)=0,\\
    -1 & X(t)>0.
    \end{cases}
\end{align*}
We can also compactly express the queuing dynamics with an infinitesimal generator $\mathcal L$ such that for any function $g:\mathbb R_{\ge0}\times\{1,2\}\times \mathcal S\to\mathbb R_{\ge0}$ that is smooth in the continuous argument.\\

If $x \ge s - \theta_1$, which means that when the current vehicle arrives, the cool-down time between this vehicle and the previous one is not over, then the arrived vehicle needs to wait for $\big(x-(s-\theta)\big)$, where $\theta = \theta_1$, if these two vehicles are in the same direction; $\theta = \theta_2$, if these two vehicles are in distinct directions. By summing up the waiting time and the crossing time $s'$, we can update $x'$ as $x' = x+s'+\theta-s$:
\begin{align*}
    &\mathcal Lg(x,y,s)=
     -g'(x,y,s)\\
     &\quad+\lambda_y\Big(\sum_{s'\in\mathcal S}p_{s'} g(x+s'+\theta_1-s,y,s')-g(x,y,s)\Big)\\
     &\quad+\lambda_{-y}\Big(\sum_{s'\in\mathcal S}p_{s'} g(x+s'+\theta_2-s,-y,s')-g(x,y,s)\Big).
\end{align*}

If $s -\theta_2 \le x < s -\theta_1$, which means that if the arrived vehicle is in the same direction as the previous one, then it can cross the intersection without waiting, otherwise it still need to wait for $\big(x-(s-\theta_2)\big)$. By summing up the waiting time and crossing time, we can update $x'$:
\begin{align*}
    & \mathcal Lg(x,y,s)= -g'(x,y,s)\\
    &\quad +\lambda_y\Big(\sum_{s'\in\mathcal S}p_{s'} g(s',y,s')-g(x,y,s)\Big)\\
    &\quad +\lambda_{-y}\Big(\sum_{s'\in\mathcal S}p_{s'} g(x+s'+\theta_2-s,-y,s')-g(x,y,s)\Big).
\end{align*}

If $0 \le x < s -\theta_2$, which means that when the current vehicle arrives, the cool-down time between this vehicle and the previous one is over, and the arrived vehicle can cross the intersection without waiting. Therefore $x'$ is updated as $x' = s'$:
\begin{align*}
    & \mathcal Lg(x,y,s)= -\mathbb{I}_{x>0}g'(x,y,s)\\
    &\quad +\lambda_y\Big(\sum_{s'\in\mathcal S}p_{s'} g(s',y,s')-g(x,y,s)\Big)\\
    &\quad +\lambda_{-y}\Big(\sum_{s'\in\mathcal S}p_{s'} g(s',-y,s')-g(x,y,s)\Big).
\end{align*}

For the above equations, $\lambda_{-y}$ denotes $\lambda_k$ such that $k\in\{1,2\}$ and  $k\ne y$, and $s$ denotes the crossing time of the last arrived vehicle, while $s'$ denotes the crossing time of the current arriving vehicle.

\section{Throughput and delay}
\label{sec:stability}

In this section, we quantify the throughput of signal-free intersections by studying stability of the queuing model.
We define throughput based on the notion of stability.
We say that the queuing system is \emph{stable} if there exists $Z<\infty$ such that for any initial condition $(x,y,s)\in\mathbb R_{\ge0}\times\{1,2\}\times \mathcal S$, $\overline D(x,y,z) \le Z$, where $\overline D(x,y,z)$ denotes the average time delay.

The main result of this paper states when such an upper bound $Z$ exists and if yes, how to quantify it.

\begin{Theorem}
Consider an intersection with arrival rates $\lambda_1,\lambda_2$ in two directions, mean crossing time $\bar S$, offset time $\theta_1$, and switch-over time $\theta_2$.
Then, the queues are stable under the FCFS sequencing policy if
\begin{align}
    \max\{\lambda_1,\lambda_2\}(\theta_2-\theta_1)+(\lambda_1+\lambda_2)(\theta_1+\overline s -s_{\min})<1.
\label{equ:stable criteria}
\end{align}
Furthermore, if the above holds, then the average time delay $\bar D$, which is the actual average system time minus the theoretical minimum system time, is upper-bounded by
\begin{align}
    \bar D\le \frac{\frac12(\lambda_1+\lambda_2)\sum_{s'\in \mathcal S}p_{s'}s'^2}{1-(\lambda_1+\lambda_2)(\theta_1+\overline s - s_{\min}) - \max\{\lambda_1, \lambda_2\}(\theta_2 - \theta_1)}.
\label{equ:upper bound}
\end{align}
\end{Theorem}

\begin{figure}[htbp]
	\centering%
	\begin{minipage}{1\linewidth}
		\centering
		\subcaptionbox{CAV intersections with 100 time steps\label{fig:CAV_num100_heatmap}}
		{\includegraphics[width= 8cm]{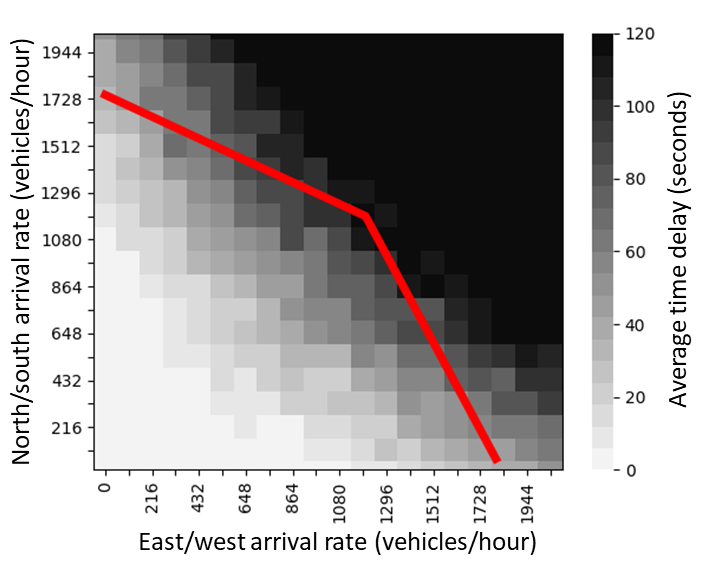}}%
	\end{minipage}
	\begin{minipage}{1\linewidth}
	    \centering
		\subcaptionbox{conventional intersections with 100 time steps\label{fig:conventional_num100_heatmap}}
		{\includegraphics[width= 8cm]{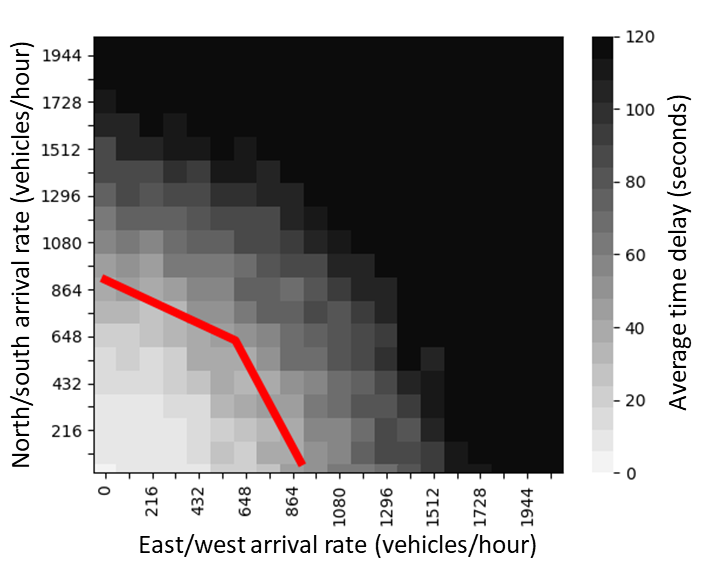}}%
	\end{minipage}
	\caption{heatmaps of CAV signal-free intersections and conventional signal-free intersections}
	\label{fig:heatmap}
\end{figure}

\begin{figure}[htbp]
    \centering
    \includegraphics[width = 8cm]{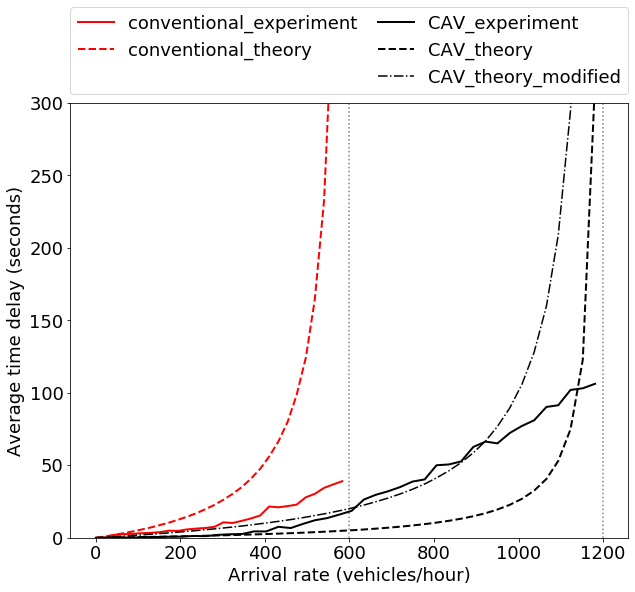}
    \caption{line plot of average time delay}
    \label{fig:lineplot}
\end{figure}

We can present the stability criteria (\ref{equ:stable criteria}) by plotting the border lines, where $\max\{\lambda_1,\lambda_2\}(\theta_2-\theta_1)+(\lambda_1+\lambda_2)(\theta_1+\overline s -s_{\min}) = 1$. As shown in figure \ref{fig:heatmap}, the red lines in the heat maps represent the border lines, and the area under the red lines satisfies our stability criteria. The upper bound of average time delay can be presented in a 2-D line plot by letting $\lambda_1 = \lambda_2$. As shown in figure \ref{fig:lineplot}, the red dashed line represents the theoretical upper bound of the conventional signal-free intersections, and the gray dashed line represents that of the CAV signal-free intersections, with equal arrival rate in each direction. Since all these figures combines the theoretical results with the experiment results, we will illustrate them in details in section \ref{sec:simulation}.

The main proof idea is that we first construct the function $g(x,y,s)$ as a quadratic Lyapunov function $V(x,y,s) = \frac12|x|^2$. The infinitesimal generator $\mathcal L$ is piece-wise according to $x$, and it becomes linear for $x \ge s_{\max} - \theta_1$. Hence a negative slope $-c < 0$ will ensure that the Lyapunov function drift in the negative direction towards zero to achieve system stability. Finally, for $0 \le x < s_{\max} - \theta_1$, we prove the piece-wise function is bounded by $d$. In all, we prove that $\mathcal LV(x,y)\le-cx+d, \forall (x,y,s)\in\mathcal X\times\mathcal Y\times\mathcal S$, and according to \cite[Theorem 4.3]{meyn1993stability} the average time delay $\overline D$ is upper-bounded by $d/c$. The detailed proof is attached below.

\noindent\emph{Proof.}
Consider the Lyapunov function
\begin{align*}
    V(x,y,s)=\frac12|x|^2.
\end{align*}
Then we have:\\
If $x \ge s-\theta_1$:
\begin{align*}
& \mathcal LV(x,y,s) = \Big(-1+\lambda_y(\theta_1+\overline s -s)+\lambda_{-y}(\theta_2+\overline s -s)\Big)x\\
    & \quad + \frac12\lambda_y\sum_{s'\in\mathcal S}p_s(\theta_1+\overline s -s)^2
    +\frac12\lambda_{-y}\sum_{s'\in\mathcal S}p_s(\theta_2+\overline s -s)^2,
\end{align*}
If $s - \theta_2 \le x < s-\theta_1$:
\begin{align*}
& \mathcal LV(x,y,s) = \Big(-1+\lambda_{-y}(\theta_2+\overline s -s)\Big)x - \frac12 \lambda_{y} x^2\\
    &\quad + \frac12\lambda_y\sum_{s' \in \mathcal S}p_{s'}s'^2 + \frac12\lambda_{-y}\sum_{s' \in \mathcal S}p_{s'}(\theta_2 + s' - s)^2,
\end{align*}
If $0 \le x < s - \theta_2$:
\begin{align*}
\mathcal LV(x,y,s) = -x - \frac12 (\lambda_{y} + \lambda_{-y}) \Big(x^2 - \sum_{s' \in \mathcal S}p_{s'}s'^2\Big).
\end{align*}

For $x\ge s_{\max}-\theta_1$, to ensure the stability, the function $\mathcal LV(x,y,s)$ should decrease as $x$ increases. Therefore, we have
$$\lambda_y(\theta_1+\overline s -s)+\lambda_{-y}(\theta_2+\overline s -s) < 1, \forall s\in \mathcal S.
$$

For $s - \theta_2 \le x < s-\theta_1$, we can compute the vertex $x^*$:
\begin{align*}
    x^* & = \frac{-1+\lambda_{-y}(\theta_2 + \overline s - s)}{\lambda_y} < \frac{-\lambda_y(\theta_1+\overline s -s)}{\lambda_y} \\
    & = s - \theta_1 - \overline s < s - \theta_2.
\end{align*}

Therefore, $\mathcal LV(x,y,s)$ is still decreasing for $s - \theta_2 \le x < s-\theta_1$.

For $0 \le x < s - \theta_2$, similarly, we can compute the vertex $x^*$:
$$x^* = \frac{-1}{\lambda_y+\lambda_{-y}} < 0.$$
Therefore, $\mathcal LV(x,y,s)$ is still decreasing for $0 \le x < s - \theta_2$. For $x < s_{\max} - \theta_1$, $\mathcal LV(x,y,s)$ is decreasing, and the maximum is obtained at $x = 0$:
$$\mathcal LV_{\max} = \frac12(\lambda_{y} + \lambda_{-y})\sum_{s' \in \mathcal S}p_{s'}s'^2.$$

Therefore we have
$$\lambda_y(\theta_1+\overline s -s_{\min})+\lambda_{-y}(\theta_2+\overline s -s_{\min}) < 1,
$$
$$
\mathcal LV(x,y)\le-cx+d,
\quad\forall (x,y,s)\in\mathcal X\times\mathcal Y\times\mathcal S,
$$
where $c = 1 - \lambda_y(\theta_1+\overline s -s_{\min}) - \lambda_{-y}(\theta_2+\overline s -s_{\min}), d = \frac12(\lambda_{y} + \lambda_{-y})\sum_{s' \in \mathcal S}p_{s'}s'^2$. Hence, by \cite[Theorem 4.3]{meyn1993stability} the average time delay $\overline D$ is upper-bounded by \eqref{equ:upper bound}.
\hfill$\square$D 
\section{Simulation-based validation}
\label{sec:simulation}

To validate the above results, we use Simulation of Urban Mobility (SUMO \cite{krajzewicz2010traffic}) to simulate the time delay for each vehicle due to the congestion. The network consists of a two-lane east/west road crossing a two-lane north/south road in a single intersection. Vehicles appear on the inbound roads with given arrival rates, pass through the central intersection, and then exit the system on the outbound roads. The vehicle flow from west to east merges with the flow from south to north with no turning allowed, and the merging sequence is determined according to the FCFS rule. The vehicle attributes, such as length and width, are presented in TABLE~\ref{tab:vehicle experiment}. Based on the same network and vehicle attributes, two types of intersections, conventional signal-free intersections and CAV signal-free intersections, are used to validate our queuing model and stability criteria under distinct pairs of arrival rates. According to the simulation results, we find that 100 time steps of vehicle generation is able to distinguish a stable system and an unstable one. For each pair of arrival rates and intersection type, 20 replications can give a good estimation to the average time delay for each vehicle. And the time step size of our simulation is 0.1 second, which means the system state will be updated every 0.1 second.

\begin{table}[htbp]
  \centering
  \caption{Vehicle attributes and experiment settings.}
    \begin{tabular}{lccccc}
    \hline
          & \multicolumn{1}{l}{Length} & \multicolumn{1}{l}{Width} & \multicolumn{1}{p{3em}}{Maximal\par speed} & \multicolumn{1}{p{5em}}{Maximal\par acceleration} & \multicolumn{1}{p{5em}}{Maximal deceleration} \\
    \hline
    Vehicle & \multicolumn{1}{c}{5m} & \multicolumn{1}{c}{1.8m} & \multicolumn{1}{c}{7m/s} & \multicolumn{1}{c}{0.8m/$\text{s}^2$} & 4.5m/$\text{s}^2$    \\
    \hline
        & \multicolumn{3}{l}{Crossing distance} &  &  \\
    \hline
        Intersection & \multicolumn{2}{c}{14.4m} & &\\
    \hline
          & \multicolumn{2}{p{5em}}{Number of replications} & \multicolumn{2}{p{8em}}{Number of steps generating vehicles} & \multicolumn{1}{l}{Step size} \\
    \hline
    Experiment & \multicolumn{2}{c}{20} & \multicolumn{2}{c}{100} & 0.1s     \\
    \hline
    \end{tabular}%
  \label{tab:vehicle experiment}%
\end{table}%

\begin{figure}[htbp]
    \centering
    \includegraphics[width = 8cm]{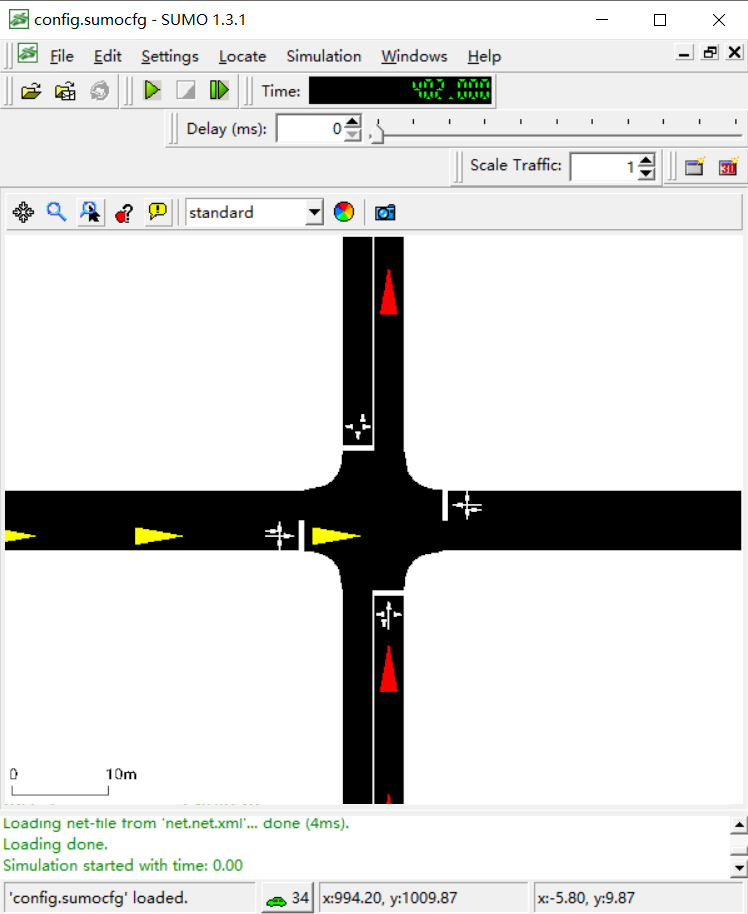}
    \caption{SUMO simulation screen shot}
\end{figure}

For ease of presentation, we assume that for the same intersection type, each vehicle has the equal intersection crossing time $s$, with ignoring the velocity fluctuation when vehicles cross the intersections.
For conventional signal-free intersections, vehicles have to come to a full stop before crossing. Hence, the crossing time is given by
\begin{align*}
    s=\sqrt{\frac{2(b+l)}a},
\end{align*}
where $a$ is the maximal acceleration, $b$ is the vehicle length, and $l$ is the crossing distance of the intersection.
In addition, large headways are needed for human-driven vehicles, which are 7.5 meters in our simulations.
For CAV signal-free intersections, vehicles can maintain the maximal speed as they cross the intersection. Hence, the crossing time is given by
\begin{align*}
    s=\frac{l+b}{\bar v},
\end{align*}
where $\bar v$ is the maximal speed allowed for crossing.
Smaller headways are allowed for CAVs. In this section, we assume that inter-CAV headways are 5.5 meters. Note that to avoid potential longitudinal collision, we need to ensure that $\overline v^2/2d \le h$, where $h$ denotes the safety headways, and $d$ denotes the maximal deceleration. And we assume that the offset time and switch-over time between two consecutive human-driven vehicles are twice as long as those between two CAVs. The following table summarizes the difference of parameter settings between theses two types of intersections.

\begin{table}[htbp]
  \centering
  \caption{Parameters for conventional vehicles and CAVs}
    \begin{tabular}{lcccc}
    \hline
          & \multicolumn{1}{p{4em}}{Minimal headway} & \multicolumn{1}{p{4em}}{Crossing time} & \multicolumn{1}{p{4em}}{Offset time} & \multicolumn{1}{p{5em}}{Switch-over time}\\
    \hline
    Conventional & 7.5m & 6.96s & 2s & 4s\\
    CAV & 5.5m & 2.77s & 1s & 2s\\
    \hline
    \end{tabular}%
  \label{tab:parameter settings}%
\end{table}%

We visualize our simulation results of the conventional signal-free intersections and the CAV signal-free intersections on two separated heat maps in Fig.~\ref{fig:heatmap}, where the shade of the cell represents the average time delay. We also plot the derived stable criteria \ref{equ:stable criteria} in the graph as the red lines, and the area under the lines satisfies our stable criteria. We believe if the average time delay exceeds 120 seconds, the system is not stable, which corresponds to the black area in the heat maps.

According to the heat maps in Fig.~\ref{fig:heatmap}, most of the area bounded by our stable criteria ~\ref{equ:stable criteria} has the light color, which represents system stability. Intuitively, we can see that the white area of the CAV intersections is larger than that of the conventional intersections. And under the same pair of arrival rates, the average system time of the CAV signal-free intersections is less than that of the conventional signal-free intersections, since the CAV has the smaller cooldown time and headways for two consecutive vehicles. Moreover, the average system time for the conventional signal-free intersections expands faster with the increase of the arrival rates.

The upper bound of average time delay under the same arrival rate is presented in Fig.~\ref{fig:lineplot}, with red lines related with the conventional intersections and black lines related with the CAV intersections. In the line plot, the dashed lines represent the theoretical upper bound of the average time delay, the solid lines represent the actual average time delay according to the simulation results, and the gray vertical dotted lines represent the stability criteria~\ref{equ:stable criteria}. We can observe that for the conventional intersections, the theoretical upper bound holds well. However, for the CAV intersections, the actual average time delay will exceed the theoretical upper bound with the increase of the arrival rate, which does not match our expectation. With tracking the vehicle crossing speed, we find that CAVs can not maintain the maximal speed when crossing the intersection under a dense arrival, due to the vehicle interaction in SUMO. Therefore, we modify the original theoretical upper bound of the CAV intersections by using the average crossing time under 0.20 vehicle arrival rate, and the average crossing time is given by the simulation results. The modified upper bound is presented in the plot by the black dash-dotted line. After modification, the theoretical line performs better.

\section{Concluding remarks}
\label{sec:conclusion}

In this paper, we formulate a two-queue polling system with Poisson arrival process, constant crossing time depending on the vehicle class, and the cooldown time between two consecutive vehicles. Vehicles cross the intersection following FCFS rules. Unlike common queuing model, we define the system state as a tuple of the residual system time, vehicle class, and crossing time of the last entered vehicle. With the state transition functions, we express the queuing dynamics with an infinitesimal generator. We further provide the analytical system stability criteria and an upper bound of the average time delay by using Foster-Lyapunov criteria. To validate our theoretical results, we use SUMO to run simulations of two signal-free intersection types. According to the simulation results, the stability criteria holds for both types of intersections, but the theoretical upper bound does not matches our expectation for the CAV intersection, which is due to the crossing time fluctuation under intersection congestion.

We provide a novel modeling method and an efficient analytical framework for vehicle sequencing at signal-free intersections. One major future work is integrating more coordination strategies into our model, with more complex state transition functions and infinitesimal generators, and we will be able to find the optimal control to minimize the average system delay. 

\bibliographystyle{plain}
\bibliography{bib_LJ,bib_Dai}   

\begin{thebibliography}{10}

\bibitem{athans1969}
Michael Athans.
\newblock a unified approach to the vehicle-merging problem.
\newblock {\em Transportation Research}, pages 123--133, 1969.

\bibitem{colombo2015}
Alessandro Colombo and Domitilla~Del Vecchio.
\newblock Least restrictive supervisors for intersection collision avoidance: A
  scheduling approach.
\newblock {\em IEEE Transactions on Automatic Control}, pages 1515--1527, 2015.

\bibitem{davis1984piecewise}
Mark~HA Davis.
\newblock Piecewise-deterministic markov processes: a general class of
  non-diffusion stochastic models.
\newblock {\em Journal of the Royal Statistical Society: Series B
  (Methodological)}, 46(3):353--376, 1984.

\bibitem{dresner2004}
Kurt Dresner and Peter Stone.
\newblock multiagent traffic management a reservation-based intersection
  control mechanism.
\newblock {\em The Third International Joint Conference on Autonomous Agents
  and Multiagent Systems}, pages 530--537, 2004.

\bibitem{guo2019}
Qiangqiang~Guo Guo, Li~Li, and Xuegang Ban.
\newblock Urban traffic signal control with connected and automated vehicles: A
  survey.
\newblock {\em Transportation Research Part C}, pages 313--334, 2019.

\bibitem{kamal2013}
Md. Abdus~Samad Kamal, Masakazu Mukai, Junichi Murata, and Taketoshi Kawabe.
\newblock Model predictive control of vehicles on urban roads for improved fuel
  economy.
\newblock {\em IEEE Transactions on Control Systems Technology}, pages
  831--841, 2013.

\bibitem{krajzewicz2010traffic}
Daniel Krajzewicz.
\newblock Traffic simulation with sumo--simulation of urban mobility.
\newblock In {\em Fundamentals of traffic simulation}, pages 269--293.
  Springer, 2010.

\bibitem{malikopoulos2014}
Andreas~A. Malikopoulos.
\newblock Supervisory power management control algorithms for hybrid electric
  vehicles: A survey.
\newblock {\em IEEE Transactions on Intelligent Transportation Systems}, pages
  1869--1885, 2014.

\bibitem{meyn1993stability}
Sean~P Meyn and Richard~L Tweedie.
\newblock Stability of markovian processes iii: Foster-lyapunov criteria for
  continuous-time processes.
\newblock {\em Advances in Applied Probability}, pages 518--548, 1993.

\bibitem{miculescu2019polling}
David Miculescu and Sertac Karaman.
\newblock Polling-systems-based autonomous vehicle coordination in traffic
  intersections with no traffic signals.
\newblock {\em IEEE Transactions on Automatic Control}, 2019.

\bibitem{ntousakis2016}
Ioannis~A. Ntousakis, Ioannis~K. Nikolos, and Morkos Papageorgiou.
\newblock optimal vehicle trajectory planning in the context of cooperative
  merging on highways.
\newblock {\em Transportation Reasearch Part C}, 2016.

\bibitem{rios2016}
Jackeline Rios-Torres and Andreas~A. Malikopoulos.
\newblock Automated and cooperative vehicle merging at highway on-ramps.
\newblock {\em IEEE Transactions on Intelligent Transportation Systems}, pages
  1--10, 2016.

\bibitem{rios2015}
Jackeline Rios-Torres, Andreas~A. Malikopoulos, and Pierluigi Pisu.
\newblock Online optimal control of connected vehicles for efficient traffic
  flow at merging roads.
\newblock {\em 2015 IEEE 18th International Conference on Intelligent
  Transportation Systems}, pages 2432--2437, 2015.

\bibitem{zhang2017b}
Yue~J. Zhang, Christos~G. Cassandras, and Andreas~A. Malikopoulos.
\newblock Optimal control of connected automated vehicles at urban traffic
  intersections: A feasibility enforcement analysis.
\newblock {\em 2017 American Control Conference (ACC)}, pages 3548--3553, 2017.

\bibitem{zhang2016}
Yue~J. Zhang, Andreas~A. Malikopoulos, and Christos~G. Cassandras.
\newblock Optimal control and coordination of connected and automated vehicles
  at urban traffic intersections.
\newblock {\em 2016 American Control Conference (ACC)}, pages 6227--6232, 2016.

\bibitem{zhang2017a}
Yue~J. Zhang, Andreas~A. Malikopoulos, and Christos~G. Cassandras.
\newblock Decentralized optimal control for connected automated vehicles at
  intersections including left and right turns.
\newblock {\em 2017 IEEE 56th Annual Conference on Decision and Control (CDC)},
  pages 4428--4433, 2017.

\bibitem{zhu2015}
Feng Zhu and Satish~V. Ukkusuri.
\newblock A linear programming formulation for autonomous intersection control
  within a dynamic traffic assignment and connected vehicle environment.
\newblock {\em Transportation Research Part C-Emerging Technologies}, pages
  363--378, 2015.

\bibitem{zohdy2012}
Ismail~H. Zohdy, Raj~Kishore Kamalanathsharma, and Hesham Rakha.
\newblock Intersection management for autonomous vehicles using icacc.
\newblock {\em 2012 15th International IEEE Conference on Intelligent
  Transportation Systems}, pages 1109--1114, 2012.

\end{thebibliography}
\end{document}